# Polaron Conductivity in $\alpha$-$Fe_2O_3$ Quenched by Adsorbed $NO_2$

*Tushar K. Ghosh[1], Elvar Ö. Jónsson[1], Stephan Steinhauer[2], Panagiotis Grammatikopoulos[3], and Hannes Jónsson[1*]*

[1]*Science Institute and Faculty of Physical Sciences, Univ. of Iceland, VR-III, 107 Reykjavík, Iceland*

[2]*Department of Applied Physics, KTH Royal Institute of Technology, Albanova University Centre, Roslagstullsbacken 21, 106 91 Stockholm, Sweden*

[3]*Instituto Regional de Investigación Científica Aplicada (IRICA) & Departamento de Física, Universidad de Castilla-La Mancha, Ciudad Real, Spain*

E-mail: hj@hi.is

**Abstract**

Polaron-mediated charge transport in $\alpha$–$Fe_2O_3$ plays a central role in its performance as a gas-sensing material, yet the atomistic interaction between surface adsorbates and polarons remains insufficiently understood. Here, density functional theory with Hubbard-U correction (DFT+U) combined with nudged elastic band calculations is used to investigate polaron formation, migration, and quenching at the Fe-terminated $\alpha$–$Fe_2O_3$ (0001) surface. The calculated activation energy for small-polaron hopping in bulk $\alpha$–$Fe_2O_3$ is found to be 0.12 eV, in excellent agreement with experimental measurements, confirming the validity of the computational approach. Slab calculations show that migration of the polaron from bulk to the surface lowers the energy by 0.12 eV, indicating preferential localization of charge carriers at the gas–solid interface. Adsorption of $NO_2$ induces substantial electron transfer (0.72 $e^-$) from the oxide to the molecule, eliminating the localized $Fe^{2+}$ polaron state and thereby suppressing polaronic conductivity. These results provide a direct microscopic explanation for the resistance increases of hematite-based sensors upon exposure to oxidizing gases. More broadly, the study establishes how surface adsorption can modulate charge transport $\alpha$–$Fe_2O_3$ through control of polaron populations, offering design principles for improved iron oxide gas sensors.

## 1. Introduction

Iron oxides are widely studied transition metal oxides for sensing applications due to their chemical stability, abundance, and rich electronic properties [1-3]. Consequently, iron oxide-based sensors have been extensively explored for environmental monitoring of air pollutants such as $NO_2$, industrial process control, automotive emission monitoring, and indoor air-quality assessment. In particular, $\alpha$–$Fe_2O_3$ (hematite) has attracted significant attention as a functional material for gas sensing, photocatalysis, and energy-related applications [4-6]. Among semiconductor metal oxides, hematite offers a particularly attractive combination of low cost, earth abundance, chemical stability, environmental compatibility, and versatile surface chemistry, which can be readily tailored through nanostructuring and defect engineering to enhance sensing performance [2,3,7,8]. Its sensitivity to oxidizing and reducing gases is largely governed by surface-mediated charge transfer processes, which directly influence electrical conductivity [9]. Understanding the microscopic mechanisms underlying charge transport and its interaction with adsorbates is therefore essential for improving sensing performance.

Charge transport in $\alpha$–$Fe_2O_3$ is dominated by small polaron hopping rather than band-like conduction [10,11]. In this mechanism, charge carriers localize on Fe sites, forming $Fe^{2+}$ centers, and migrate via thermally activated hopping between neighboring lattice sites. The activation energy and mobility of these polarons are key parameters controlling the electrical response of hematite-based sensors. Previous studies have shown that polaron transport can be strongly affected by structural distortions, defect states, and surface interactions, which are all highly relevant under sensing conditions.

The role of temperature in polaron-mediated charge transport has been extensively investigated in hematite and other transition-metal oxides. In $\alpha$-$Fe_2O_3$, electrical conduction occurs predominantly through thermally activated small-polaron hopping, with both experiments and first-principles calculations demonstrating that charge carrier mobility increases with temperature [10-14]. Since metal oxide gas sensors typically operate at elevated temperatures, thermal activation directly influences electrical conductivity and sensor response. Similar behavior has been reported in, e.g., ceria-based materials, where Paul *et al.* showed that temperature governs the polaronic contribution to electrical conduction in both bulk devices and $CeO_2$ thin films [15,16]. Although ceria and hematite differ in crystal structure and defect chemistry, these studies reinforce the general importance of temperature in controlling polaron-mediated transport across transition-metal oxides.

From a sensing perspective, α–$Fe_2O_3$ has demonstrated promising performance for detection of gases such as $NO_2$, acetone, and other volatile compounds [17,18]. The sensing mechanism is typically attributed to adsorption-induced charge transfer, which modifies the concentration or mobility of charge carriers near the surface. In particular, oxidizing gases such as $NO_2$ are known to extract electrons from the oxide, thereby increasing the resistance of metal oxide semiconductors with n-type characteristics [19]. However, a detailed atomistic understanding of how such adsorption processes interact specifically with polaronic charge carriers remains incomplete.

In earlier experimental work, it was observed that iron oxide systems can initially form γ–$Fe_2O_3$ (maghemite) rather than α–$Fe_2O_3$ [18]. While γ–$Fe_2O_3$ is often metastable, it is well established that it transforms into the thermodynamically stable α–$Fe_2O_3$ phase, especially at elevated temperatures relevant for gas sensing operation [1,20]. This phase transition underscores the importance of understanding charge transport and surface interactions in α–$Fe_2O_3$, as it is the phase most likely to dominate under realistic device conditions.

Despite extensive experimental studies, there remains a need for atomistic insight into the coupling between polaron transport and gas adsorption at α–$Fe_2O_3$ surfaces. In particular, the extent to which adsorbates can modify or suppress polaron conduction is a critical question for interpreting sensing signals. Previous theoretical studies have largely addressed either small-polaron formation and transport in bulk hematite or adsorption-induced charge transfer at the surface, whereas the direct coupling between these processes has received comparatively less attention. In particular, understanding how adsorption influences the stability and mobility of surface polarons is essential for establishing a microscopic connection between charge transport and the electrical response of hematite-based gas sensors. Density functional theory (DFT) combined with harmonic transition state theory provides a powerful framework to address this issue by explicitly resolving charge localization, migration pathways, and adsorption-induced electronic changes [21].

In this work, we investigate polaron formation and migration in bulk α–$Fe_2O_3$ and at the Fe-terminated (0001) surface using DFT+U calculations. The activation energy for polaron hopping is determined using the nudged elastic band method. Furthermore, the interaction between surface polarons and $NO_2$ molecules is analyzed to elucidate the microscopic origin of sensing behavior. We therefore examine the interaction of $NO_2$ with the surface polaron and demonstrate that adsorption induces substantial charge transfer from the hematite surface to the adsorbed molecule, thereby quenching the localized polaronic state. Since small-polaron hopping constitutes an important electronic transport mechanism in hematite, this adsorption-induced polaron quenching provides a microscopic mechanism by which $NO_2$ adsorption can suppress electronic conduction and consequently increase the electrical resistance. The results provide direct insight into how adsorbates can quench polaron conductivity, thereby establishing a clear link between atomistic charge transport and macroscopic sensing response.

## 2. Methods and Model

The calculations were performed using density functional theory (DFT) employing the VASP software [22,23] using the PBE exchange correlation functional [24] as well as the PBE+U approach. The valence electrons were described using a plane wave basis with a 500 eV cutoff while the effect of inner electrons was described using the PAW formalism. The addition of a Hubbard U term helps reduce the effect of the self-interaction error, which artificially increases the energy of localized electronic states with respect to localized states; without this correction, polaronic states are unstable. We chose the value U = 4.3 eV as in earlier calculations of α–$Fe_2O_3$ [12]. In the structural relaxations, the energy was optimized with respect to atomic positions until the forces on the ions in the bulk unit cells were converged to 0.01 eV/Å. The criterion for convergence of the self-consistent charge density was a tolerance on the energy change of $10^{-5}$ eV.

Bulk properties of α–$Fe_2O_3$ were calculated using a $Fe_{12}O_{18}$ unit cell with 4×4×2 k-point sampling. In order to find the optimized lattice parameter, the equilibrium cell shape for different volumes was calculated; the lattice parameters were extracted from the equilibrium cell shape that corresponds to the lowest energy.

A conventional periodic slab model was used for simulating the Fe-terminated (0001) surface. A slab of 2×2×1 $Fe_{48}O_{72}$ supercell was used for (0001) α–$Fe_2O_3$ surface. Supercells were generated using the optimized lattice parameters of the bulk α–$Fe_2O_3$. In order to ensure that the periodic images do not interact, the box length along the c-direction was set to ensure at least 12 Å of vacuum between the slab and its periodic images, and the k-point sampling was reduced to 2×2×1. $Fe_{48}O_{72}$ slabs were chosen in such a way that the top and the bottom layers were equivalent and symmetric with respect to the center of inversion located at the center of the slab. The symmetry prevented unphysical dipolar interaction between periodic images of the slab. The choice of such a model was essential for achieving well-converged slab properties. For instance, the calculated surface energy obtained with this model is 1.8 J $m^{-2}$, which falls within the range of 1.08-1.9 J m $^{-2}$ reported in previous DFT studies [25,26], demonstrating good agreement with the available literature.

A polaron was formed by adding an excess electron in a 2×2×1 super-cell, $Fe_{48}O_{72}$, with 2×2×2 k-point sampling. The excess negative charge was compensated by a uniform positive background. Both the bulk and slab calculations were performed with the same net charge and using an identical computational setup. This consistent treatment ensures that the dominant electrostatic contribution arising from the charge largely cancels when evaluating the relative stabilization energy of the polaron. In this way, the need to determine absolute charged-defect formation energies is avoided and instead focuses on the relative energies of the excess electron. Consequently, image-charge corrections, which are generally required for obtaining accurate absolute formation energies of charged defects in periodic supercells, are not necessary for the relative energy comparisons considered here.

The nudged elastic band (NEB) method [27,28] with the climbing image extension [29] was used to find the activation energy for polaron migration. Five intermediate images were generated between the initial and final minimum energy structures to provide a discrete representation of the path. Adjacent images were connected by spring forces with a spring constant of 5.0 eV Å$^{-2}$, while the climbing-image algorithm was employed to accurately locate the saddle point configuration of the atoms. The electronic self-consistent calculations were converged to an energy tolerance of $10^{-5}$ eV, and the atomic coordinates were considered converged when the maximum magnitude of the Hellmann–Feynman force acting on each atom was less than 0.01 eV Å$^{-2}$.

## 3. Results

The calculated bulk atomic structure and the orientation of the magnetic moments of the Fe atoms is shown in **Figure 1**a. In order to find the optimal lattice parameter, the lowest energy cell shape was identified. Calculated lattice parameters (Figure 1b) are tabulated in **Table I** and compared with experimental values as well as PBE values without the U term. The agreement with experiment is good, within 1%.

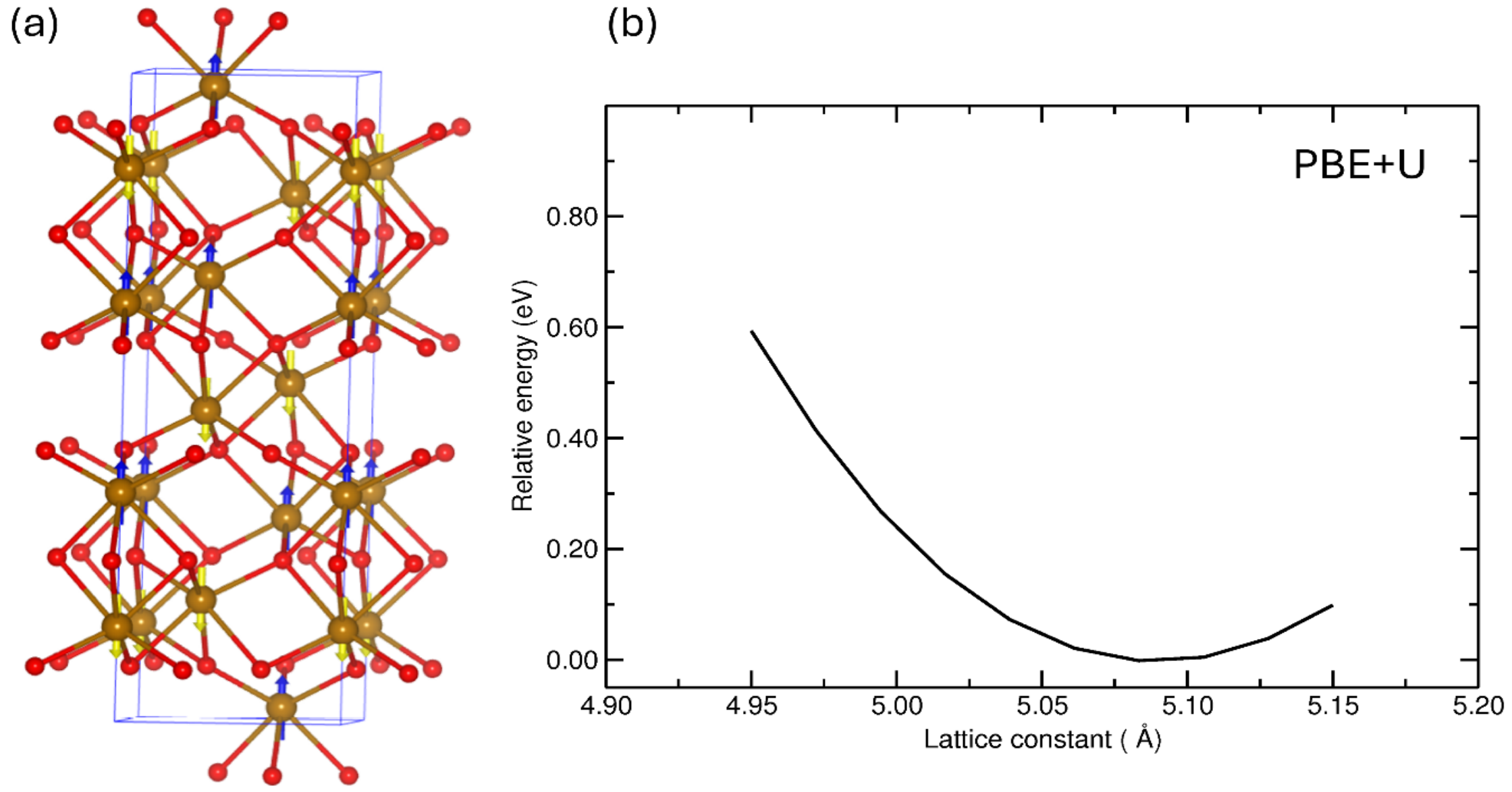

**Figure 1. (a)** Atomic structure of the $Fe_{12}O_{18}$ unit cell of the crystal and corresponding orientation of the magnetic moments. Brown and red spheres represent Fe and O atoms, respectively. **(b)** Energy of α-$Fe_2O_3$ with respect to volume change monitored as the change of the lattice constant a.

**Table I.** Calculated and experimental cell parameters of α-$Fe_2O_3$ bulk using a unit cell of $Fe_{12}O_{18}$.

| Structural Parameter | PBE Calculation | PBE+U (U=4.3) Calculation | Experiments [30] |
|---|---|---|---|
| a (Å) | 5.02 | 5.08 | 5.04 |
| b (Å) | 5.02 | 5.08 | 5.04 |
| c (Å) | 13.91 | 13.90 | 13.77 |
| α (deg) | 90.0 | 90.0 | 90.0 |
| β (deg) | 90.0 | 90.0 | 90.0 |
| γ (deg) | 120.0 | 120.0 | 120.0 |

The iron atoms in α-$Fe_2O_3$ were found to be ferromagnetically coupled within the basal plane but were antiferromagnetically coupled along the c-axis, as shown in Figure 1. All the iron atoms in α-$Fe_2O_3$ were in the +3 oxidation state. The calculated magnetic moment of a $Fe^{3+}$ ions was 4.18 $\mu_B$, similar to previous calculations at the same level of theory [12]. This also compares well with the experimental value of 4.6 $\mu_B$. However, calculations using the PBE functional without the U term gave a magnetic moment of only 3.60 $\mu_B$.

In order to compute the ground state properties of α-$Fe_2O_3$ in the presence of a polaron, an excess electron was added to a 2×2×1 α-$Fe_2O_3$ supercell and the size of the simulation box was optimized again. The structure was fully relaxed to compute lattice properties in presence of the excess charge; the optimized lattice parameters are tabulated in **Table II**. The presence of the polaron had negligible effect on the simulation box size, indicating that the latter was large enough. Results of calculations using the PBE functional without the U term are also shown in Table II for comparison. The excess electron was localized at an iron site and reduced one of the $Fe^{3+}$ ions to an $Fe^{2+}$ ion. The electron density difference is shown in **Figure 2**, where the presence of the excess electron breaks the symmetry of the lattice. It is important to emphasize that standard PBE does not stabilize a localized small-polaron state because of the inherent self-interaction error. Instead, the excess electron remains delocalized over the system (see Figure 2c-d). The excess electron can be localized only by introducing a Hubbard U term, which helps to reduce the self-interaction error and thereby enables polaron formation. The distortion of the bond lengths is observed near the $Fe^{2+}$ ion, where the polaron is localized. The angles of the simulation cell distort slightly in the presence of the polaron, as listed in Table II.

**Table II.** Calculated cell parameters of 2×2×1 α-$Fe_2O_3$ in the presence of a polaron in the system, calculated using both PBE and PBE+U.

| Structural Parameter | PBE Calculation | PBE+U (U=4.3) Calculation |
|---|---|---|
| a (Å) | 10.04 | 10.15 |
| b (Å) | 10.04 | 10.15 |
| c (Å) | 13.93 | 13.90 |
| α (deg) | 90.04 | 90.10 |
| β (deg) | 89.96 | 90.00 |
| γ (deg) | 119.95 | 120.00 |

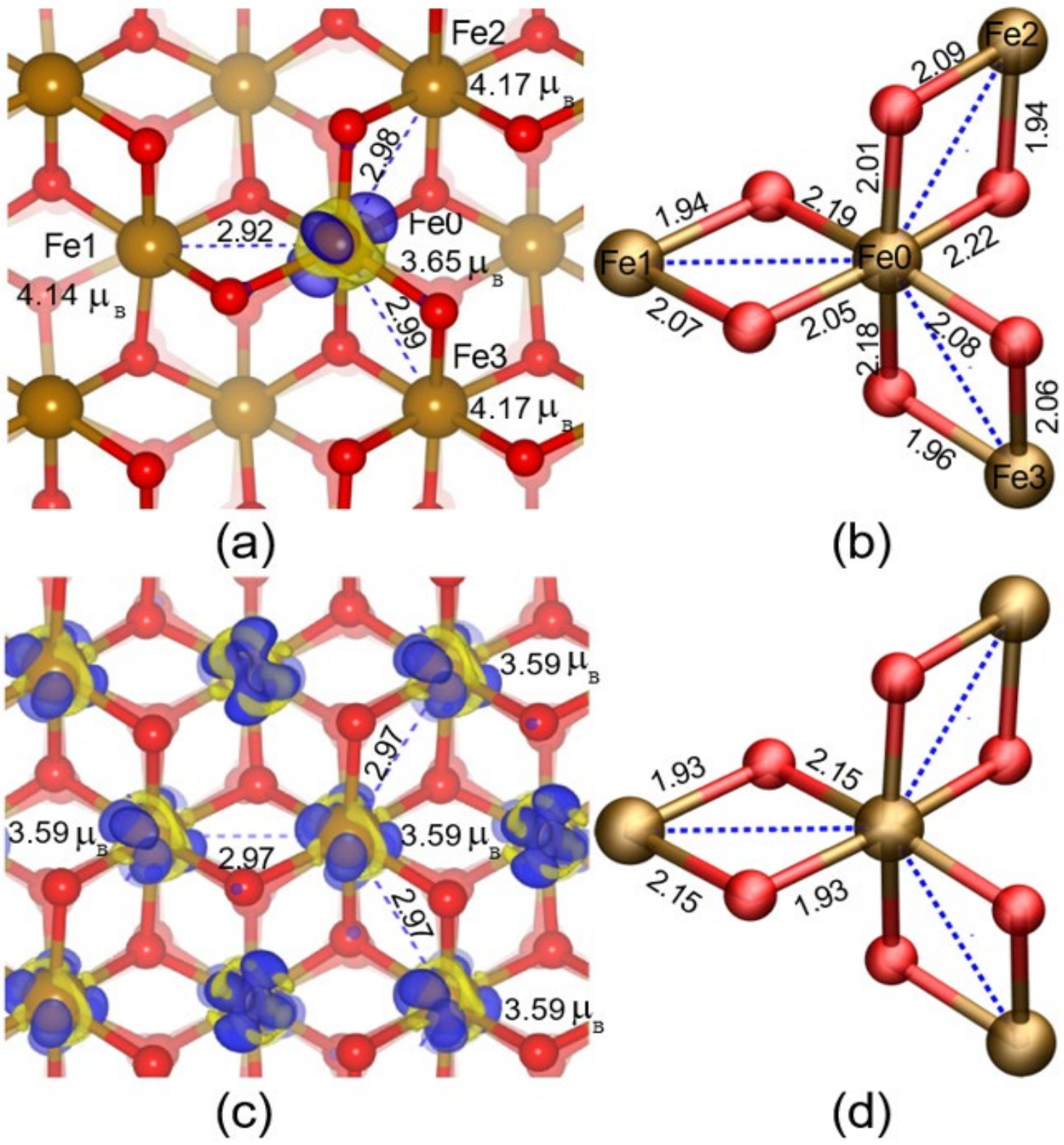

**Figure 2.** Atomic structure and electron density difference of the polaron in α-$Fe_2O_3$ calculated using PBE+U (**(a)** and **(b)**), and using PBE without the U term (**(c)** and **(d)**). **(a)** and **(c)**: 2×2×1 supercell with an excess electron viewed along the c-axis. The electron density difference, $\rho_{\text{diff}}^{\text{bulk}} = \rho_{\overline{\text{Fe}}_2\text{O}_3} - \rho_{\text{Fe}_2\text{O}_3}$, is shown, where blue indicates increased density and yellow indicates decreased density (where the overscore identifies the structure with the polaron present). Electron densities are illustrated as the isosurfaces with ±0.1*e*. The excess electron is located at the atom labeled Fe0. The distances to other iron atoms in the same basal plane, labeled Fe1, Fe2 and Fe3, are marked and indicated with blue dashed line. **(b)** and **(d)** Length of bonds between Fe and O atoms near the $Fe^{2+}$ ion.

The minimum energy path for the migration of the polaron from one Fe ion to an adjacent one within the basal plane was calculated using the NEB method. The results are shown in **Figure 3**. The rise in the energy along the minimum energy path gives an activation energy of 0.12 eV. This is in excellent agreement with the experimental estimate of 0.118 eV [13].

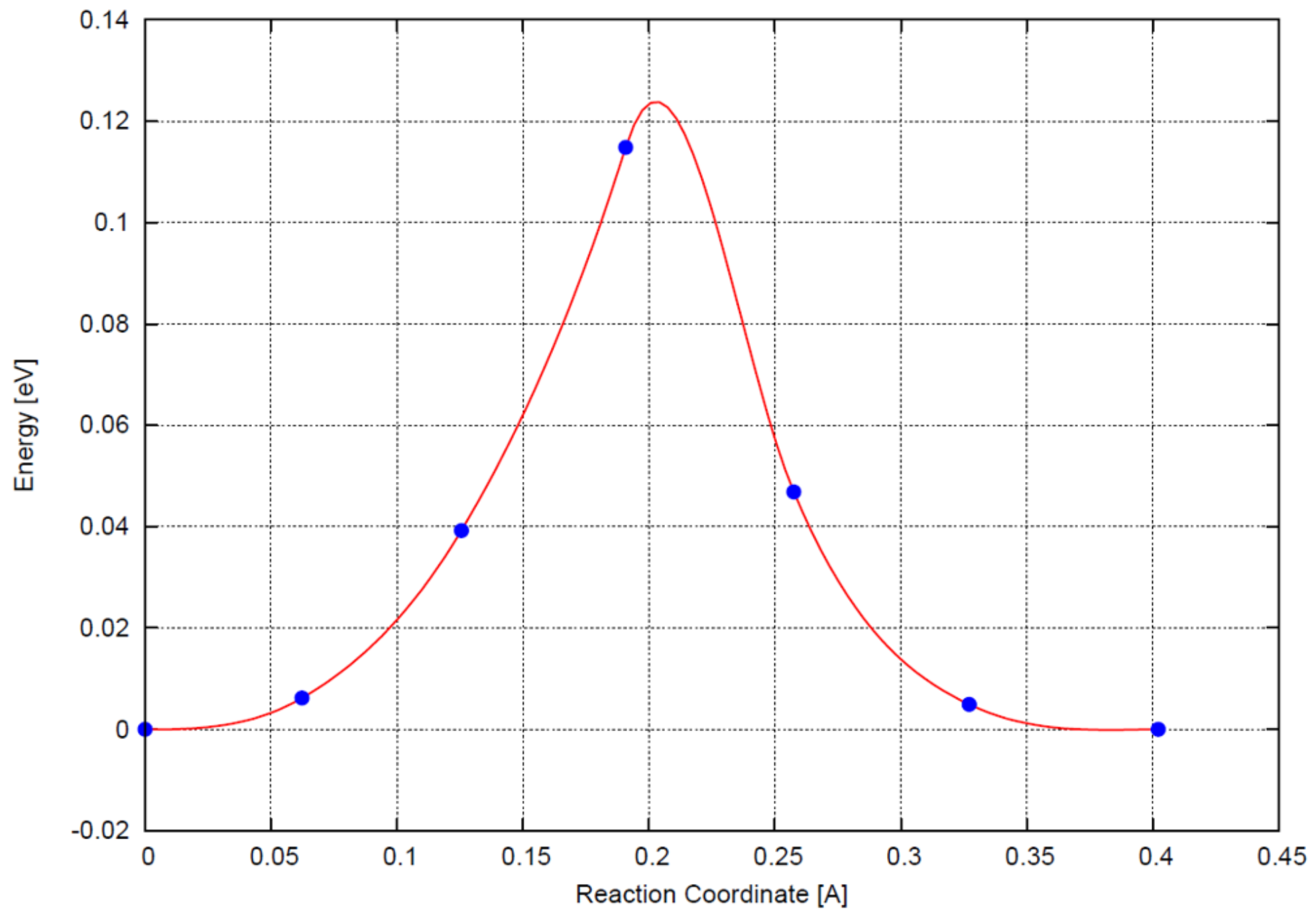

**Figure 3.** Minimum energy path found using NEB calculations for polaron migration between two adjacent Fe ions, Fe0 and Fe1, within the same basal plane (see Figure 2a) in the crystal.

An excess electron was then added to the slab to generate a polaron at the Fe terminated (0001) surface. The energy of the polaron is lower at the surface than in the interior of the crystal by 0.12 eV. A polaron will, therefore, tend to migrate to the surface due to energetic preference, but entropic effects will counter that to some extent due to the large number of sites within the crystal compared with the number of surface sites (**Figure 4**a). A $NO_2$ molecule was then adsorbed on the (0001) surface at the site of the polaron (Figure 4b). The calculated adsorption energy of $NO_2$ on the Fe-terminated (0001) surface in the presence of a localized surface polaron is −2.16 eV. This indicates a relatively strong adsorption and is consistent with the substantial charge transfer from the surface to the $NO_2$ molecule observed in the Bader charge analysis. The electron transfer, as determined by Bader analysis, is 0.72 electrons (see **Table III**). The presence of the admolecule therefore removes the polaron from the α-$Fe_2O_3$. This would largely eliminate the electrical conductivity due to presence of polarons preferentially located at surface sites.

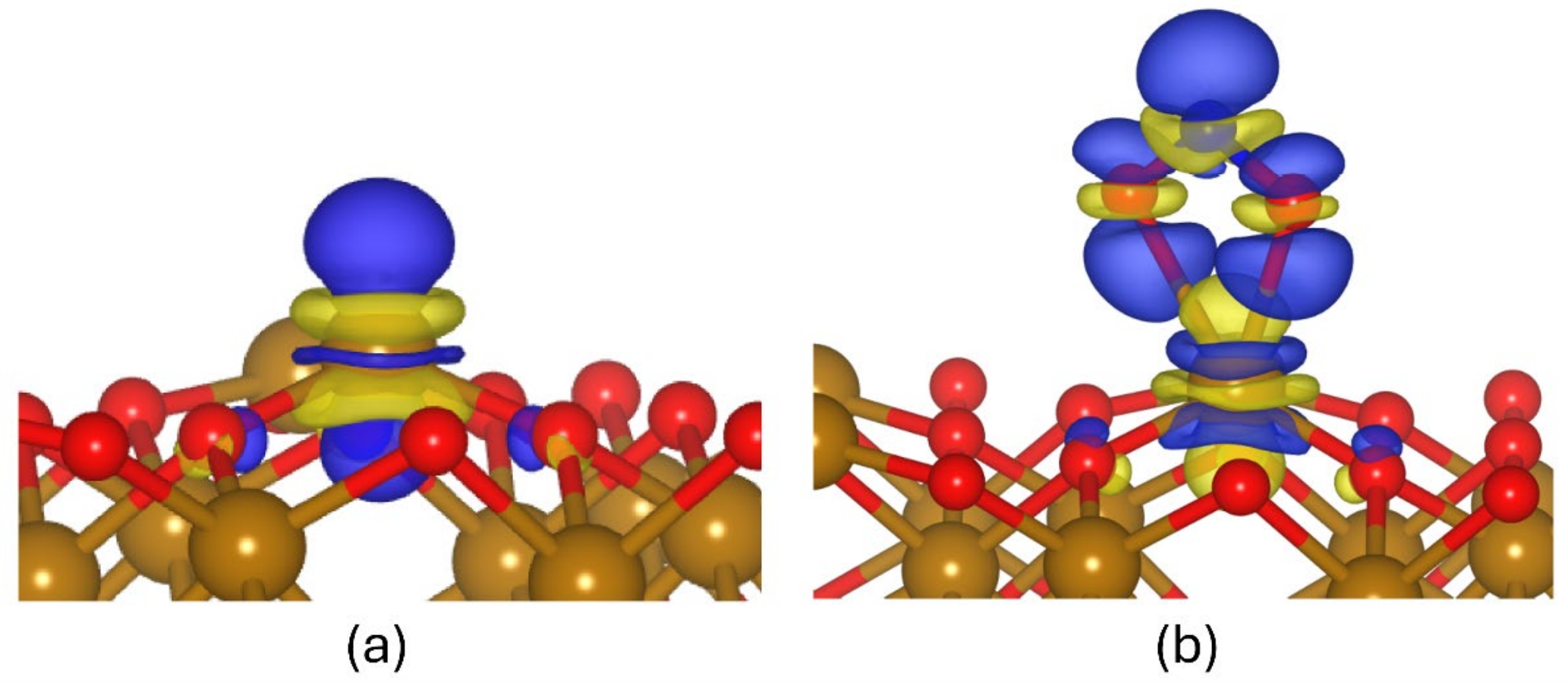

**Figure 4. (a)** Electron density difference, $\rho_{\text{diff}}^{\text{surf}} = \rho_{\overline{\text{Fe}}_2\text{O}_3} - \rho_{\text{Fe}_2\text{O}_3}$, between a charged and a neutral (0001)-$Fe_2O_3$ surface, due to the presence of excess electron at the Fe-terminated (0001) surface. **(b)** Electron density difference, $\rho_{\text{diff}}^{\text{NO}_2} = \rho_{\overline{\text{Fe}}_2\text{O}_3} - \rho_{\text{Fe}_2\text{O}_3}$, between a $NO_2$-adsorbed (0001)-α-$Fe_2O_3$ surface with a polaron and a clean (0001)-α-$Fe_2O_3$ surface with a polaron plus an isolated $NO_2$ molecule. Electron densities are illustrated as the isosurfaces with ±0.1*e*. Blue isosurfaces indicate a net gain of electrons, whereas yellow isosurfaces indicate a net loss of electrons. Fe, O, and N atoms are depicted brown, red, and blue, respectively.

**Table III.** Change in electron density integrated over Bader volumes due to adsorption of $NO_2$ on Fe-terminated (0001)–2×2×1 *α*-$Fe_2O_3$ surface with a polaron. The total negative charge on the $NO_2$ is 0.72 electrons.

| Electron gain by $NO_2$ | | | Electron loss by Fe | | | |
|---|---|---|---|---|---|---|
| O1 | N | O2 | Fe0 | Fe1 | Fe2 | Fe3 |
| 0.32 | 0.12 | 0.28 | 0.48 | 0.07 | 0.06 | 0.07 |

## 4. Discussion

The calculated activation energy for polaron migration (0.12 eV) is in close agreement with the experimental estimate, confirming that the DFT+U framework accurately captures the small-polaron transport mechanism in α–$Fe_2O_3$ [10,11,21]. This low activation barrier indicates that charge transport is feasible at moderate temperatures, consistent with the operating conditions of many gas sensors based on hematite. From a sensing standpoint, this implies that the electrical conductivity is highly sensitive to perturbations that affect either the polaron population or their mobility. We point out, however, that the PBE+U approximation with U = 4.2 eV gives an underestimation of the band gap by about 0.5 eV. A better approach, but also significantly more computationally intensive, would be to apply explicit self-interaction correction [31,32] or use a hybrid functional where screened Fock exchange is mixed in with the PBE functional [33].

A key finding of this work is the energetic preference of polarons for the surface relative to the bulk. The stabilization of the polaron by 0.12 eV at the Fe-terminated (0001) surface suggests that charge carriers naturally accumulate near the surface region, where gas adsorption occurs. This is particularly important for sensing applications, as it implies that the active charge carriers are inherently localized at the interface where interaction with analyte molecules takes place. Such surface localization enhances the sensitivity of the material, since even small perturbations in surface chemistry can significantly influence the overall conductivity.

The adsorption of $NO_2$ has a pronounced effect on the electronic structure of the system. Bader charge analysis [34] shows substantial electron transfer (0.72 $e^-$) from the oxide to the adsorbed molecule. This charge transfer effectively removes the localized electron responsible for polaron formation, thereby eliminating the corresponding $Fe^{2+}$ site. As a result, the polaron conduction pathway is disrupted. This mechanism provides a clear atomistic explanation for the experimentally observed increase in resistance upon exposure to oxidizing gases such as $NO_2$.

Importantly, the quenching of polaron conductivity is not merely a local effect but can have a global impact on transport. Since conduction in α–$Fe_2O_3$ occurs via hopping between localized states, the removal of even a fraction of these states can significantly reduce the connectivity of the conduction network. In systems where polarons preferentially reside at the surface, adsorption-induced charge transfer can therefore lead to a strong modulation of electrical resistance, forming the basis of the sensing signal. Incidentally, if $NO_2$ initially adsorbs at a neighboring $Fe^{3+}$ site, the low polaron migration barrier (0.12 eV) suggests that the surface polaron may readily migrate toward the adsorption site. The resulting adsorption-induced charge transfer could stabilize the adsorbed $NO_2$ molecule while simultaneously quenching the localized polaronic state.

These findings also highlight the importance of surface structure and termination. The Fe-terminated (0001) surface stabilizes polarons and facilitates strong interaction with $NO_2$. Variations in surface morphology, defect density, or crystallographic orientation could therefore lead to substantial differences in sensing performance. This is consistent with experimental observations that nanostructured α–$Fe_2O_3$ (e.g., nanotubes, nanocubes) exhibits enhanced sensitivity due to increased surface area [3,17,18,35].

Furthermore, the results provide insight into the role of phase stability. While γ–$Fe_2O_3$ may initially form in some synthesis routes (such as gas-phase synthesis of nanoparticles [18,36]), its metastability and transformation to α–$Fe_2O_3$ at elevated temperatures imply that the sensing behavior in practical devices will largely be governed by hematite [37], unless it can be ensured that the sensing material remains maghemite during the sensor lifetime. The polaron-based mechanism identified here is therefore directly relevant for interpreting sensing responses under realistic operating conditions.

While the present study establishes how surface adsorption can modulate charge transport in α–$Fe_2O_3$ through polaron migration, thereby providing microscopic insight relevant to the design of improved iron-oxide-based gas sensors, several additional factors may influence the sensing response under realistic operating conditions. In particular, surface hydroxylation, intrinsic defects, oxygen vacancies, adsorbed oxygen species, humidity, grain boundaries, operating temperature, and the exposed crystallographic facet can significantly affect both adsorption energetics and polaron-mediated charge transport. The presence of such structural and environmental complexities could alter polaron formation, localization, and migration, as well as the interaction between polarons and adsorbed gas molecules. A systematic investigation incorporating these effects, particularly the role of different surface defects and adsorbed species, therefore represents a logical and valuable extension of the present work toward a more comprehensive understanding of charge transport and gas sensing in hematite under realistic conditions.

## 5. Conclusion

In this work, density functional theory calculations have been used to investigate polaron formation, migration, and interaction with adsorbates in α–$Fe_2O_3$. The calculated activation energy for polaron hopping is in excellent agreement with experimental measurements, confirming the validity of the small-polaron transport model in this material.

A key result is the preferential stabilization of polarons at the Fe-terminated (0001) surface, which implies that charge carriers are naturally localized in the region where gas adsorption occurs. This surface localization enhances the sensitivity of α–$Fe_2O_3$ to adsorbates. The adsorption of oxidizing gases such as $NO_2$ leads to significant electron transfer from the oxide to the molecule, effectively removing the polaron and quenching polaron-mediated conductivity.

This mechanism provides a clear microscopic explanation for the sensing behavior of α–$Fe_2O_3$ toward oxidizing gases, where adsorption results in increased electrical resistance. The results emphasize the critical role of polaron dynamics in determining sensor response and highlight the importance of surface structure in optimizing sensing performance. To this end, further studies are needed to compare the effect of different surface orientations from a theoretical perspective. While the present calculations establish the energetic preference for polaron localization at the surface relative to the bulk, a detailed investigation of subsurface localization, inclusion of defects, hydroxylation, adsorbed oxygen, humidity, and grain boundaries, temperatures remain an important direction for future work.

Overall, the findings establish a direct link between atomistic charge transport processes and macroscopic sensing properties in α–$Fe_2O_3$. This understanding can guide the design of improved iron oxide-based sensors through control of surface structure, defect engineering, and phase composition.

## Acknowledgments

This work was supported by the Icelandic Research Fund (grant no. 2410644). P.G. was supported by a Beatriz Galindo Senior Fellowship (BG23/00144).